\documentclass[prl,twocolumn,floatfix,showpacs,superscriptaddress]{revtex4}
\usepackage{amsmath,amsfonts}
\usepackage{epsfig}

\newcommand{\la}{\lambda}
\newcommand{\bea}{\begin{eqnarray}}
\newcommand{\beq}{\begin{equation}}
\newcommand{\eea}{\end{eqnarray}}
\newcommand{\eeq}{\end{equation}}
\def\simgeq{\; 
\raisebox{-0.4ex}{\tiny$\stackrel
{{\textstyle>}}{\sim}$}\;}

\begin{document}
\title
{Large $N$ Scaling Behavior of the Lipkin-Meshkov-Glick Model}
\date{\today}
\author{F.~Leyvraz}
\affiliation{Centro de 
Ciencias F\'\i sicas, Universidad Nacional
Aut\'onoma de Mexico, 
Cuernavaca, 62551 Morelos, Mexico}
\author{W.D.~Heiss}
\affiliation{Institute of Theoretical Physics and Department of 
Physics, University of Stellenbosch, 7602 Matieland, South Africa }
\begin{abstract}
We introduce a novel semiclassical approach to the Lipkin model. In
this way the well-known phase transition arising at the critical value 
of the coupling is intuitively understood. New results --
showing for strong couplings the existence of a threshold energy which separates
deformed from undeformed states as well as the divergence of 
the density of states at the threshold energy -- 
are explained straightforwardly and in quantitative terms 
by the appearance of a double well structure in a
classical system corresponding to the Lipkin model. Previously
unnoticed features of the eigenstates near the threshold energy are
also predicted and found to hold.
\end{abstract}
\pacs{03.65.Sq, 05.30.Fk, 64.60.-i}
\maketitle
The Lipkin-Meshkov-Glick (LMG) model, originally introduced in nuclear
physics has found applications in a broad range of other topics: 
statistical mechanics of quantum spin systems \cite{bot82},
Bose--Einstein condensates \cite{cir98} as well as quantum
entanglement \cite{vid04}, to name but a few. The continued interest 
in this system arises from the fact that it is an exactly solvable
\cite{pan99, lin03} many-body interacting quantum system as well as
one of the simplest to 
show a quantum transition in the regime of strong coupling. 

This
transition is by now well-understood: the ground state becomes
degenerate and a macroscopic change in the ground state energy takes
place. Furthermore, at the transition value of the coupling, the
density of states diverges at the ground state energy as the number
$N$ of interacting particles becomes large. The energy gap
was recently found \cite{duvi} to vanish as $N^{-1/3}$. Furthermore,
in this model a novel type of phase transition
has recently been discovered for strong  values of the 
coupling parameter. Indeed, in this regime the spectrum is divided 
by a {\em critical energy} $E_c$, where the behavior characteristic of 
strong coupling holds below $E_c$, while above $E_c$
the system reverts to the kind of behavior found below the phase
transition. At the critical energy, the density of states
is again found to diverge as the particle number $N$ goes to infinity. 
This divergence has been conjectured to be of the logarithmic type.

In recent papers 
\cite{duvi,hsg} different approaches have been used to explain these 
results. 
The continuous unitary
transformation technique (flow equations) \cite{duvi,kms} 
was applied to obtain the spectrum reliably
for large $N$ values. An investigation 
of the singularities of the 
spectrum \cite{hsg}
in the complex $\la$-plane \cite{hesa} (the exceptional points 
\cite{kato}) sheds more
light upon the complexity of the limit problem but a final answer about 
the limit attained has not been
given. These singularities have been 
recognized as an essential 
mechanism to
invoke the phase transition in a 
seminal paper by Lee and Yang \cite{leey} and their significance for 
the partition function is discussed more recently again in 
\cite{borrm,cej}.
A different approach \cite{duvi} starts with the bosonization 
method using higher orders of powers in $1/N$ of the 
Holstein-Primakoff representation and then 
applies the flow equation technique. New 
results have been obtained in this way, in particular 
the correct analytic 
behavior of the level distance as a function of $N$ at the 
critical point. Earlier attempts \cite{hemu} have revealed 
different results such as
the form of the wave functions beyond the phase transition.

In this Letter we introduce a novel semiclassical approach to the
LMG model. It readily explains the above
features, determines the precise value of the critical energy as a
function of the coupling, proves the logarithmic divergence of the
density of states near the critical energy as well as successfully
predicts certain previously unnoticed 
behavior of the eigenstates near the critical energy.
Finally, we obtain a qualitative understanding of all essentials of the model:
the classical model we introduce has a double well
structure above the phase transition, and the critical energy can then
be identified with the separatrix energy. The
approach given here also shows easily 
both the nature of the phase transition as a function of the
coupling parameter as well as the scaling with  $N$ of the vanishing
gap at the critical coupling, which was previously shown \cite{duvi}
to scale as $N^{-1/3}$. 

We here recapitulate the basics of the 
model and discuss the essential properties for large 
values of $N$. It is given in terms of 
$2j+1=N+1$-dimensional representations of
the SU(2) operators $J_k,\,k=x,y,z$ as follows
\beq
H(\la 
)=J_z+\frac{\la}{N}(J_x^2-J_y^2).
\label{lip}
\eeq
Here the interaction is scaled by $N$ to ensure that 
$H$ is extensive. In this 
form the model has a phase transition just beyond 
$\la =1$, the larger $N$
the closer the transition point at $\la =1$. 
This has been discussed under various points 
of view in the literature, see 
e.g.~\cite{pan99,ring}.

The Hamiltonian allows reduction into two spaces: 
$m$ integer and $m$ half-integer, with $m$ the eigenvalues 
of $J_z$; it corresponds to $N$ even and odd respectively
and is denoted as parity. For $\la 
\simgeq 0$ the even and odd levels
are obviously separated and 
remain so for all $\la <1$ while the levels become 
degenerate (up to 
terms vanishing exponentially fast in $N$) 
for $\la >1$. The phase at $\la<1$ is called the normal phase,
while the symmetry (parity)
breaking phase at $\la >1$ is called the 
deformed phase. Recent calculations 
\cite{kms} apply non-perturbative flow 
equations allowing  to obtain 
the spectrum for arbitrarily 
high yet finite values of $N$. These have 
established \cite{hsg} the existence of the phase transition in energy
referred to above: the states having energy below a certain threshold 
behave as states of the deformed phase, whereas higher in the spectrum
the states become undeformed again. 

Since the commutator 
$[H(\la ),(\vec J)^2]$ vanishes, we confine 
ourselves to a fixed
value of 
$j=N/2$. For large $N$ we consider the 
Hamiltonian (\ref{lip}) on the 
sphere of radius $j=N/2$. In other words, we 
rewrite the Hamilton 
operator (\ref{lip}) as a classical Hamilton function
\beq
H=\frac{N}{2}\bigg(
    -\sin \theta \cos 
\phi-\frac{\lambda}{2}
(\cos^2\theta-\sin^2 \theta \sin^2\phi )
    \bigg)
\label{hpol}
\eeq
having introduced the polar angles as \cite{foot}
\bea
J_z & = & 
-\frac{N}{2}\sin \theta 
\cos\phi  \cr
J_x & = & \frac{N}{2}\cos \theta  
\cr
J_y & = & 
\frac{N}{2}\sin \theta \sin \phi .  \cr
\nonumber
\eea
Note that the transition to a classical Hamiltonian has also been
achieved in a different way using the coherent state approach developed 
in \cite{sch98}.
At this point we notice that, with $\mu =\cos \theta$, 
the Poisson bracket
\beq
\{\mu,\phi 
\}=\frac{2}{N}
\label{poiss}
\eeq
suggests how to quantize
the Hamilton function of the
single particle problem in the two
canonical conjugate coordinates $\phi$ 
and $\mu$. It can be written as
\beq
K\equiv\frac{2H}{N}=-\sqrt{1-\mu 
^2}\cos\phi-\frac{\lambda}{2} 
(\mu^2-(1-\mu^2)\sin^2\phi 
).
\label{sp}
\eeq
To obtain information about the ground state and the low lying 
states we expand $2H/N$ around its minimum which is found 
at
\bea
\sin 
\phi _0&=& 0 
\label{eq:phi-min}\\
\mu _0 &=& 
\left\{
\begin{array}{lr}
0 & (\la\leq1)  \\
    \pm\sqrt{1-\la ^{-2}} 
&(\la\geq1)
\end{array}
\right.
\label{eq:mu-min}
\eea
and the corresponding minimum values of $H$ at
\beq
H= \left\{
\begin{array}{lr}
-\frac{N}{2}&(\la\leq 
1)\\
-\frac{N}{4}(\la + \la^{-1}) &(\la \ge 
1)
\end{array}
\right.
\label{minval}
\eeq
Expanding around the minimum reveals in fact all 
essential features. Keeping to lowest
order terms we obtain around 
$\mu=\phi=0$
\beq
K=-1+\frac{1+\la}{2}\phi ^2 +\frac{1-\la}{2} 
\mu^2+
\frac{\mu^4}{8}+\ldots
\label{exp}
\eeq
In this form, quantization is 
straightforward. Based on 
(\ref{poiss}) we identify
$\phi $ and $\mu $ 
with 
momentum and position, respectively, and $2/N$ 
with $\hbar $, 
i.e.~we 
use
the usual canonical commutation relations for $\mu$ and $\phi$.
With this identification the 
Hamiltonian (\ref{exp}) represents a 
quartic oscillator which behaves
for $\la <1$ basically like a traditional oscillator with a harmonic 
spectrum (for the lower states)
$E_k\sim k \sqrt{1-\la 
^2},\,k=1,2,\ldots $. For $\la>1$, the lower
states must be determined at the minimum 
in $\mu $ around the values $\mu _0$ 
as given in 
(\ref{eq:mu-min}). 
With $\epsilon =\mu -\mu _0$ the expansion of the Hamiltonian 
$K$ 
yields, up to the constant given in 
(\ref{minval})
\beq
K=\frac{\la^2-1}{2}\la\epsilon^2+\frac{\phi^2}{\la}.
\label{eq:5}
\eeq
For low-lying 
levels, the harmonic spectrum is again obtained
from (\ref{eq:5}) with
frequency $\sqrt{2(\la^2-1)}$ well-known from previous work
\cite{pan99,gehe}.

To evaluate the average energy density, the features of which 
have been the object of recent work \cite{hsg}, we exploit, as we are 
in the semi-classical regime, the WKB relation
\beq
S(2E_{k}/N)=2\pi\left(k+\frac{1}{2}\right)\hbar .
    \label{eq:1}
\eeq
Here $S(E)$ denotes the action corresponding to the
Hamiltonian $K$. By differentiation we obtain
\beq
    \Delta E=E_{k+1}-E_{k}=\frac{\pi\hbar 
N}{T(2\overline E/N)}
    =\frac{2\pi}{T(\overline E)}
    \label{eq:2}
\eeq
where 
$T(E)$ is the period of the orbit with respect to $K$ as a
function of energy and $\overline E$ is $(E_{k+1}+E_{k})/2$. Use
is made of $T(E)=dS/dE$. 

The last relation (\ref{eq:2}) contains virtually all basic 
information of the Lipkin model. Firstly, for $\lambda<1$ nothing 
dramatic happens: the density of states merely changes smoothly as the 
period varies. An important fact should still be noted: the 
bosonization approach predicts, a constant 
$\Delta E$ in 
this case. We see here that this does not hold over the entire 
energy range: the period varies
smoothly over large energy scales, and 
so does the average energy spacing; it increases with the
energy. 

Secondly, for $\lambda>1$ there is an
energy, in fact the separatrix, 
being situated for $K$ at the value
$-1$, where the period $T(E)$ {\em 
diverges}. To estimate the
average energy spacing in this region, we consider
trajectories near the classical separatrix where they spend 
a long time near the
unstable equilibrium, that is, in a region in which
$\phi$ as well as $\mu$ are small.
One has for $K$ approximately
\beq
K=-1+\frac{1+\la}{2}\phi^2-\frac{\la 
-1}{2}\mu^2.
\label{eq:3}
\eeq
Since this dominates the divergence of 
$T(E)$ near $E=-1$, one finds
that $T(E)$ is approximately given by 
$\ln|2E/N+1|/\sqrt{\la^2-1}$. But
$2E/N+1$ is itself of the order $2\Delta 
E/N$ and hence of first order in $1/N$. We thus obtain
\beq
    \Delta 
E=\frac{2\pi \sqrt{\la^2-1}}{\ln N} .
    \label{eq:4}
\eeq
This corresponds to 
the high density of states 
observed in \cite{hsg} for 
specific values of 
$\la>1$. It was found that 
a change occurred between the two regimes 
below and above a certain $\la $-dependent energy 
$E_{c}(\la)$: for low energies $E<E_{c}(\la )$, the 
states were deformed, the order parameter was non-zero
and an odd-even 
degeneracy was observed. For $E>E_{c}(\la )$, all these 
phenomena disappeared and a normal regime, similar to
$\la<1$ was recovered. The 
transition region between those two regimes 
had the typical signature of high
density of states. From (\ref{eq:2}) it becomes clear 
that the lower portion of 
the two regimes correspond to bounded motion in one well 
breaking parity symmetry. We mention that the tunneling 
between the left hand and right hand
wells determines the splitting to 
be $\sim \exp 
(-{\rm const}/\hbar)\sim \exp (-{\rm const}\,N)$. 
For higher energies, the corresponding classical motion 
is above the wells and symmetry is 
restored. The two different 
regimes are separated by the separatrix with its 
high density of states. 
Formula (\ref{eq:4}) has 
been verified numerically as shown in Fig.~1. 
\begin{figure}[t]
\begin{center}
\epsfig{file=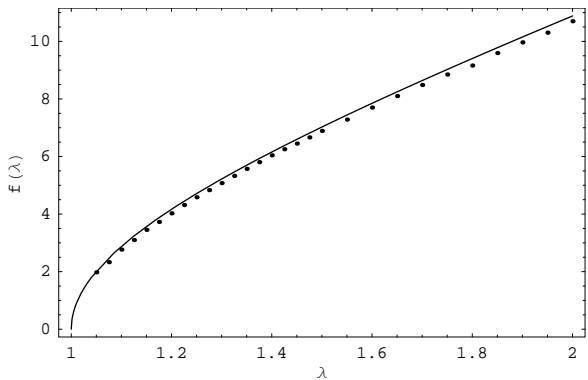,height=5.0cm,width=8.0cm,clip=,angle=0}
\caption{Asymptotic 
behavior of the distance of levels of one
parity at the transition point 
as a function of $\la $. The points are the numerical
fits for 
$500 < N <1500$ when fitted to $f(\la )/\ln(N)$; the solid 
curve is $2 \pi \sqrt{\la ^2-1}$.}
\label{asy}
\end{center}
\end{figure}
The region of high density is predicted to occur at energy $-1$ 
corresponding to energies above the ground state around 
$(\la+\la^{-1}-2)/2$ which again is verified numerically.

As the third major result, (\ref{eq:2}) provides the leading 
analytic behavior of the spectrum at the transition point ($\la=1$).
Since the Hamiltonian 
is quartic for (classically) low energies, one can 
evaluate $\Delta
E$ using the formula for $T(E)$ for a quartic oscillator 
valid for energies high up in the
spectrum but still small with respect to $N$.
It is given by $T(E)={\rm const.}\cdot E^{-1/4}$, from 
which
\beq
\Delta E={\rm const.}\cdot (E/N)^{1/4}
\label{eq:6}
\eeq
follows, and it entails
\beq
E_k\sim k^{4/3}/N^{1/3}.
\label{eq:4.3}
\eeq
The 
$N^{-1/3}$ behavior has been obtained recently \cite{duvi} and the 
$k^{4/3}$ 
behavior is confirmed numerically (see Fig.2). We stress the  
non-uniform nature of the limit played by
the critical point 
$\la =1$ when comparing (\ref{eq:6}) and (\ref{eq:4}).

It is 
obvious that 
the spectra of the two 
Hamiltonians (\ref{lip}) and 
(\ref{hpol})
differ. This is due to the issue of ordering: in order to make sense
of (\ref{hpol}), we must specify in which way we order $\mu$ and
$\phi$ to obtain a self-adjoint operator. There is no unique
prescription for this so that an unknown difference exists between the
two Hamiltonians. It is, however, known that if care is taken,
these errors are of order $\hbar^2$, that is, of order
$N^{-2}$. From this follows that we also expect the singularities 
to be different. We recall that they
are associated with the critical point and the transitions 
for $\la >1$. As the semi-classical treatment preserves
these basic features it is expected that at least
the qualitative pattern of the exceptional points remains. In fact,
there is (i) the special feature at $\la =1$
(an accumulation point for $N\to \infty $ \cite{hsg}), (ii)
a high density of EPs near to the separatrix \cite{hemu},
that is for energies around $E_{c}(\la )$,
(iii) the absence of singularities near to real values for $\la <1$, and 
for $\la >1$ for energies sufficiently distant from $E_{c}(\la )$.

Of interest in 
the semi-classical treatment is the behavior
of the wave function at 
the phase transition for $\la >1$.
In accordance with the long dwelling 
time classically at the 
saddle point, there is the phenomenon of 
super-scarring for
the wave function. In fact, this has been shown
generally 
\cite{colin} for the occurrence of such double wells.
It arises for the 
specific values of $k$ where the separatrix 
itself satisfies the WKB 
condition (\ref{eq:1}). 
The wave function then
shows a dramatic 
concentration at the saddle-point of the Hamiltonian
(or at the maximum in 
configuration space): in particular, it can be
shown that there is an 
interval, the length of which goes to zero 
with $\hbar$, that is with $1/N$, in which 
the whole wave function is 
asymptotically
concentrated. This is reflected in the eigenvectors of (\ref{lip})
associated 
with the eigenvalues at the minimal gap: in the basis
of $J_z$ they 
become relatively concentrated in the first few
components, that is around 
$\mu =0$; while the number of substantial
components increases with 
$N$, the first twenty components exhaust
the norm by about 50\% 
irrespective of $N$. This has been tested
for $\la =1.1,\,1.5$ and 2.0 where the 
transition occurs at 
$k\approx N/120,\,N/16$ and $N/8$, respectively. 
The effect is rather significant -- the more so the nearer $\la $ to 
unity -- in that the
first twenty components of the neighboring wave 
functions contribute appreciably less to the total norm.
\begin{figure}[t]
\begin{center}
\epsfig{file=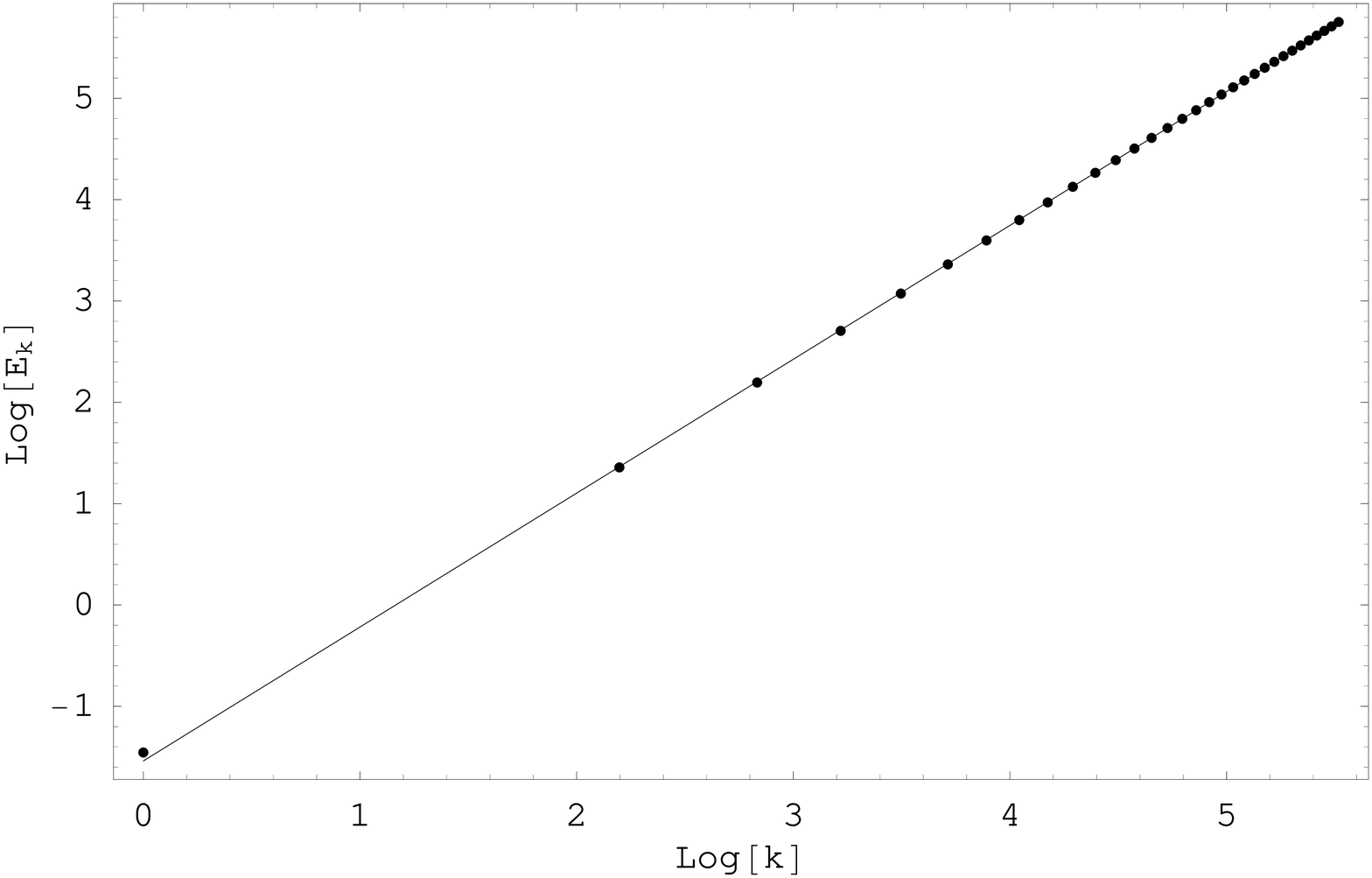,height=5.0cm,width=8.0cm,clip=,angle=0}
\caption{Log-log-plot of $E_k$ versus $k$. For clarity only every eighth point
of the first 500 levels are taken ($N=5000$). The straigth line fits
the slope 4/3 with 1\% accuracy.}
\label{k43}
\end{center}
\end{figure}
In principle, 
wave functions can be directly determined 
semi-classically. As this is 
not the major focus of this
paper we only outline the procedure. 
Switching to
the usual polar coordinates
defined around the 
$z$-axis but keeping the definition of
$\mu$ as before, the 
renormalized Hamiltonian $K$ reads
\begin{equation}
K=\mu+\frac{\lambda}{2}(1-\mu^2)\cos2\phi.
    \label{eq:7}
\end{equation}
In these 
variables, the usual spherical harmonics are eigenfunctions
of the operator $\hat\mu=(2/iN)\partial /\partial{\phi}$
with an eigenvalue proportional to 
that of $J_{z}$. 
The semi-classical eigenfunctions of 
(\ref{eq:7})
in the eigenbasis of $\hat\mu$ are then 
expressed as a function of the energy 
by means of standard WKB formulae for the one-dimensional Hamiltonian
(\ref{eq:7}).
The expressions fail, of course, in the usual manner 
near the 
turning points of (\ref{eq:7}). 

To summarize: using the semi-classical version of the 
original 
model (\ref{lip}) leads to (\ref{hpol}).The WKB approximation
then yields the expansion (\ref{exp}). 
Most information can then  be extracted from
(\ref{eq:2}) being 
based on (\ref{eq:1}). There is (i) the qualitative
result about the phase 
transition occurring for $\la >1$ as discussed
in \cite{hsg}. This 
includes the deviation from the strict
equidistant level sequence for large 
$N$ and for $\la <1$ as well as
the exponential (in $N$) separation of 
the degenerate levels for $\la >1$. There is (ii)
the expression 
(\ref{eq:4}) for the level distance at the transition point
for $\la >1$; there is (iii) the finding (\ref{eq:4.3}) at $\la =1$.
The apparent contrast of results (ii) 
and (iii) underlines once again the nonuniform nature of the
large $N$ limit at $\la =1$.
Note that all these results have been confirmed 
numerically. Additional results
referring to the semi-classical wave 
function, in particular the super-scarring
at the saddle point in phase space 
are presented. The qualitative behavior
of the singularities of the 
spectrum (the exceptional points) seems to be
preserved in the 
semi-classical approach.
\vskip 1cm
{\bf Acknowledgment} 
The authors are 
grateful to Hannes Kriel for preparation of the drawing.
One of us (WDH) 
gratefully acknowledges the generous hospitality that 
he enjoyed during 
his three weeks stay from Thomas Seligman, Director of the CIC,
and the colleagues at 
the Department of Physics at Cuernavaca, 
Mexico.

\end{document}